\documentclass{article}
\usepackage{hiph-art}
\usepackage{epsfig}
\usepackage{graphicx}
\volnumber{25} \issuenumber{1} \edyear{2006}                             
\frompage{000} \topage{000}                                              
\recrevdate{7 May 2005}                                                  
\def\rightmark{Quantum Clan Model description of BEC}
\def\leftmark{O.V.Utyuzh et al.}

\title{Quantum Clan Model description of Bose-Einstein correlations}
\authors{
{O.V.Utyuzh$^{1,a}$, G.Wilk$^{1,b}$ and Z.W\l odarczyk$^{2}$ %
\index{One, A.} 
\index{Two, A.} 
}\\[2.812mm]
{\normalsize
\hspace*{-8pt}$^1$ The Andrzej So\l tan Institute for Nuclear Studies;\\
Ho\.za 69; 00-681 Warsaw, Poland\\[0.2ex]
\hspace*{-8pt}$^2$ Institute of Physics, \'Swi\c{e}tokrzyska Academy,\\
         \'Swi\c{e}tokrzyska 15; 25-406 Kielce, Poland; E-mail:
         wlod@pu.kielce.pl\\[0.2ex]
\hspace*{-8pt}$^a$ E-mail: utyuzh@fuw.edu.pl\\[0.2ex]
\hspace*{-8pt}$^b$ E-mail: wilk@fuw.edu.pl
}}

\abstract{We propose a novel numerical method of {\it modelling}
Bose-Einstein correlations (BEC) observed among identical (bosonic)
particles produced in multiparticle production reactions. We argue that
the most natural approach is to work directly in the momentum space in
which the Bose statistics of secondaries reveals itself in their tendency
to bunch in a specific way in the available phase space. Because such
procedure is essentially identical to the clan model of multiparticle
distributions proposed some time ago, therefore we call it the {\it
Quantum Clan Model}.}

\keyword{Bose-Einstein correlations; Statistical models; Fluctuations}

\PACS{25.75.Gz 12.40.Ee 03.65.-w 05.30.Jp}

\makeindex
\begin{document}

\maketitle

\noindent
The phenomenon of Bose-Einstein correlations (BEC) is so widely
known \cite{BEC} that we shall start directly with problem of its
{\it proper numerical modelling}, such which would account from the
very beginning for the quantum statistical bosonic character of
identical secondaries produced in hadronization process. To our
knowledge this problem was so far considered only in \cite{OMT}(cf.,
however, \cite{ZAJC}). All other approaches claiming to model BEC
numerically \cite{modBEC} use as their starting point the outcomes
of existing Monte-Carlo event generators (MCG) describing
multiparticle production process \cite{GEN} and {\it modify} them in
a suitable way to fit the BEC data. These modifications are called
{\it afterburners}. They inevitably lead to such unwanted features
as violation of energy-momentum conservation or to changes in the
original multiparticle spectra. Actually, it is worth to mention at
this point that in \cite{UWW} we proposed afterburner free from such
unwanted effects. It was based on different concept of introducing
quantum mechanical (QM) effects in the otherwise purely
probabilistic distributions from those proposed in \cite{QUANT}.
Namely, each MCG provides us usually with a given number of
particles, each one endowed with one of $(+/-/0)$ charges and with
well defined spatio-temporal position and energy-momentum. On the
other hand, experiment provides us information on only the first and
last characteristics. The spatio-temporal information is not
available directly (in fact, the universal hope expressed in
\cite{BEC,modBEC} is it can be deduced from the previous two via the
measured BEC). Our reasoning was as follows: $i)$ BEC phenomenon is
of the QM origin, therefore one has to introduce in the otherwise
purely classical distributions provided by MCG a new element
mimicking QM uncertainties; $ii)$ this cannot be done with
energy-momenta because they are measured and therefore fixed; $iii)$
the next candidate, i.e. spatio-temporal characteristics, can be
changed but this was already done in \cite{QUANT,modBEC}; $iv)$ one
is thus left with charges and in \cite{UWW} we simply assigned (on
event-by-event basis) new charges to the particles selected by MCG
conserving, however, the original multiplicity of $(+/-/0)$. This
has been done in such a way as to make particles of the same charge
to be located maximally near to each other in the phase space by
exploring natural fluctuations of spatio-temporal and
energy-momentum characteristic resulting from MCG. In this way we
automatically conserve all energy-momenta and do not change
multiparticle distributions and do it already on the {\it level of
single event} provided by MCG . However, the new assignment of
charges introduces a profound change in the structure of the
original MCG. Generally speaking (cf. \cite{UWW} for details), it is
equivalent to introduce the bunching of particles of the same charge
used in the MCG.

This observation will be the cornerstone of our new proposition. Let
us remind that idea of bunching of particles as quantum statistical
(QS) effect is not new \cite{QS}. It was used in connection with BEC
for the first time in \cite{GN} and later it was a cornerstone of
the {\it clan model} of multiparticle distributions $P(n)$ leading
in natural way to their negative binomial (NB) form observed in
experiment \cite{NBD}. It was introduced in the realm of BEC in
\cite{BSWW} and \cite{OMT,ZAJC}. Because our motivation comes
basically from \cite{OMT} let us outline shortly its basic points.
It deals with the problem of how to distribute a given number of
bosonic secondaries, $\langle n\rangle =\langle n^{(+)}\rangle +
\langle n^{(-)}\rangle + \langle n^{(0)}\rangle$, $\langle
n^{(+)}\rangle =\langle n^{(-)}\rangle =\langle n^{(0)}\rangle$, in
a least biased way . Using information theory approach (cf.,
\cite{IT}) their rapidity distribution was obtained in form of grand
partition function with temperature $T$ and chemical potential
$\mu$. In addition, the rapidity space was divided into {\it cells}
of size $\delta y$ (fitted parameter) each. It turned out that
whereas the very fact of existence of such cells was enough to
obtain reasonably good multiparticle distributions, $P(n)$,
(actually, in the NB-like form), their size, $\delta y$, was crucial
for obtaining the characteristic form of the $2-$body BEC function
$C_2(Q=|p_i-p_j|)$ (peaked and greater than unity at $Q=0$ and then
decreasing in a characteristic way towards $C_2=1$ for large values
of $Q$) out of which one usually deduces the spatio-temporal
characteristics of the hadronization source \cite{BEC} (see
\cite{OMT} for more details). The outcome was obvious: to get $C_2$
peaked and greater than unity at $Q=0$ and then decreasing in a
characteristic way towards $C_2=1$ for large values of $Q$ one must
have particles located in cells in phase space which are of nonzero
size.$^a$
\begin{figure}[ht]
\hspace{1.15cm}
  \begin{minipage}[ht]{104mm}
   \insertplot{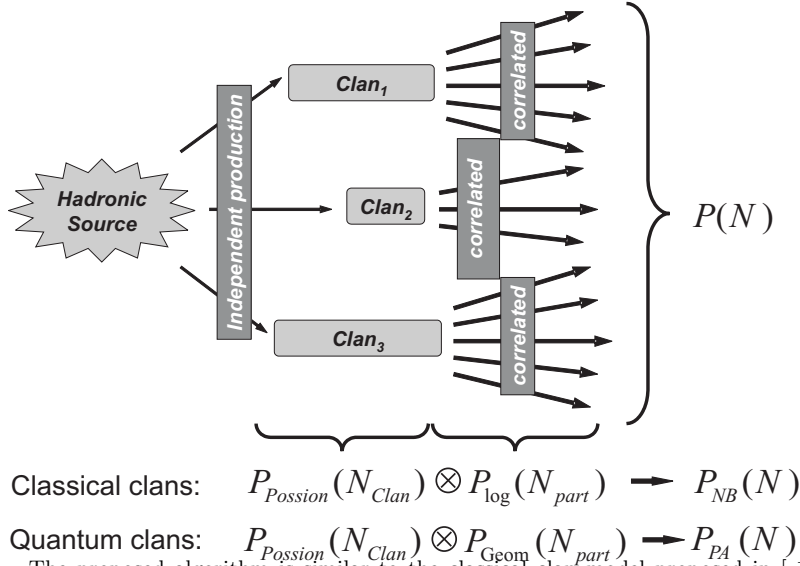}
  \end{minipage}
  \vspace{-0.5cm}
  \caption{
\footnotesize The proposed algorithm is similar to the classical clan model proposed
in \protect\cite{NBD} but its clans contain particles of the same charge and (almost)
the same energies and are distributed according to geometrical distribution what
results in overall P\`olya-Aeppli distribution, $P_{PA}(n)$ \protect\cite{PA}, insted
of NB one, $P_{NB}(n)$.}
  \label{fig:Fig1}
\end{figure}

To illustrate our proposition let us assume that mass $M$ hadronizes
into $N=\langle n\rangle$ bosonic particles (we take them as pions
of mass $m$) with equal numbers of $(+/-/0)$ charges and with
limited transverse momenta $p_T$. Suppose that their multiplicity
distribution $P(n)$ follows a NB-like form (i.e. it is broader than
Poissonian) and that their two-particle correlation function of
identical particles, $C_2(Q)$, has the specific BEC form mentioned
above. To model such process accounting from the very beginning, for
the bosonic character of produced particles we propose the following
steps (illustrated by comparison to some selected LEP $e^+e^-$ data
\cite{Data}, cf., Fig. \ref{fig:Fig1}):

{\bf 1)} Using some (assumed) function $f(E)$ select a particle of
energy $E^{(1)}_1$ and charge $Q^{(1)}$. The actual form of $f(E)$
should reflect somehow our {\it a priori} knowledge of the
particular collision process under consideration. In what follows we
shall assume that $f(E) = \exp\left( -E/T\right)$, with $T$ being
parameter (playing in our example the role of "temperature").

{\bf 2)} Treat this particle as seed of the first {\it elementary
emitting cell} (EEC) (introduced in \cite{BSWW}) and add to it,
until the first failure, other particles of the same charge
$Q^{(1)}$ selected according to distribution $P(E)=P_0\cdot
\exp\left( -E/T\right)$, where $P_0$ is another parameter (playing
the role of "chemical potential" $\mu = T\cdot \ln P_0$). This
assures that the number of particles in this EEC, $k_1$, will follow
geometrical (or Bose-Einstein) distribution, i.e. $\langle
k_1\rangle = P(E)/[1+P(E)],$ and accounts therefore for their
bosonic character. As result $C_2(Q)>1$ but only {\it at one point},
namely for $Q=0$.
\begin{figure}[ht]
  \begin{minipage}[ht]{57mm}
        \insertplot{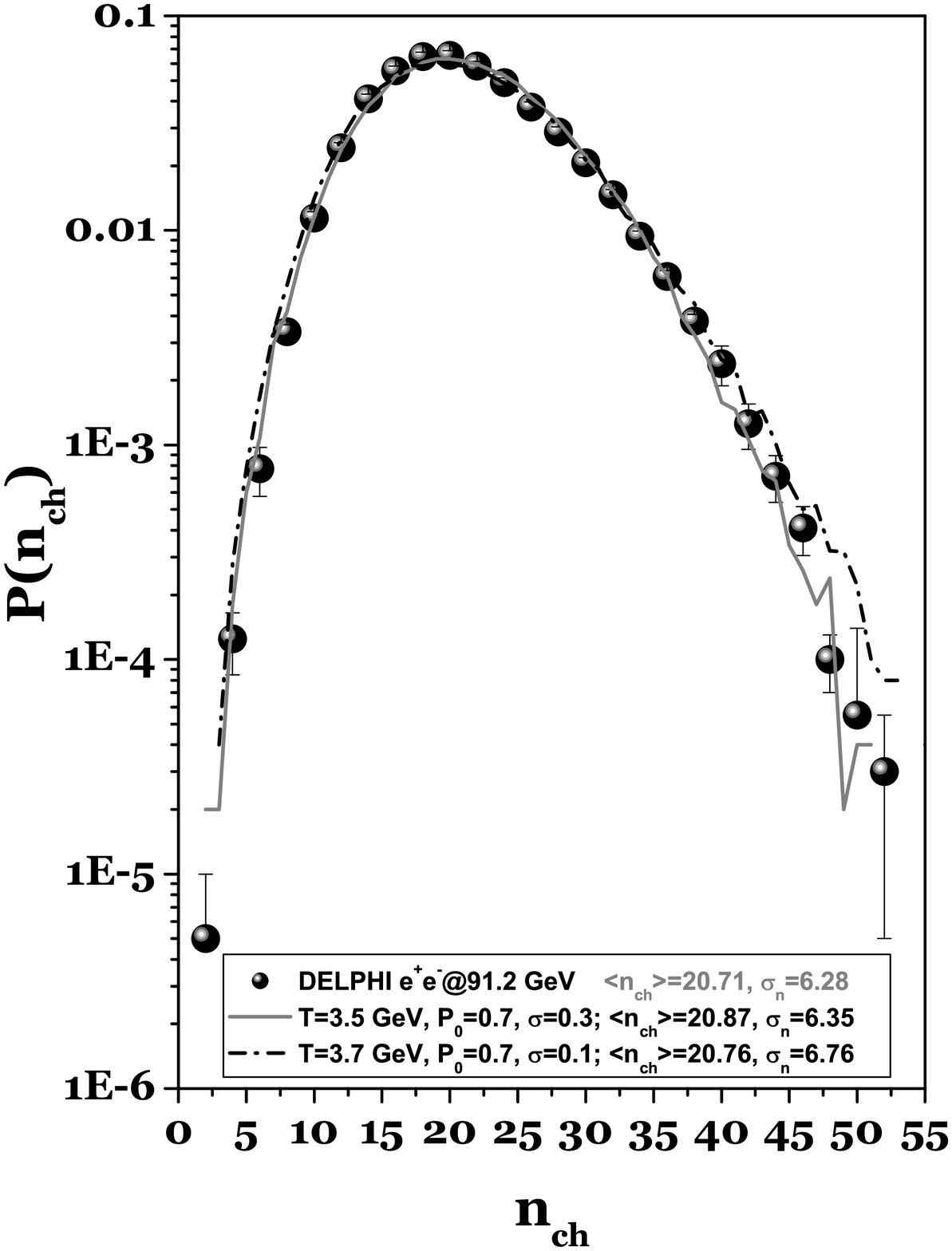}
  \end{minipage}
\hfill
  \begin{minipage}[ht]{57mm}
       \insertplot{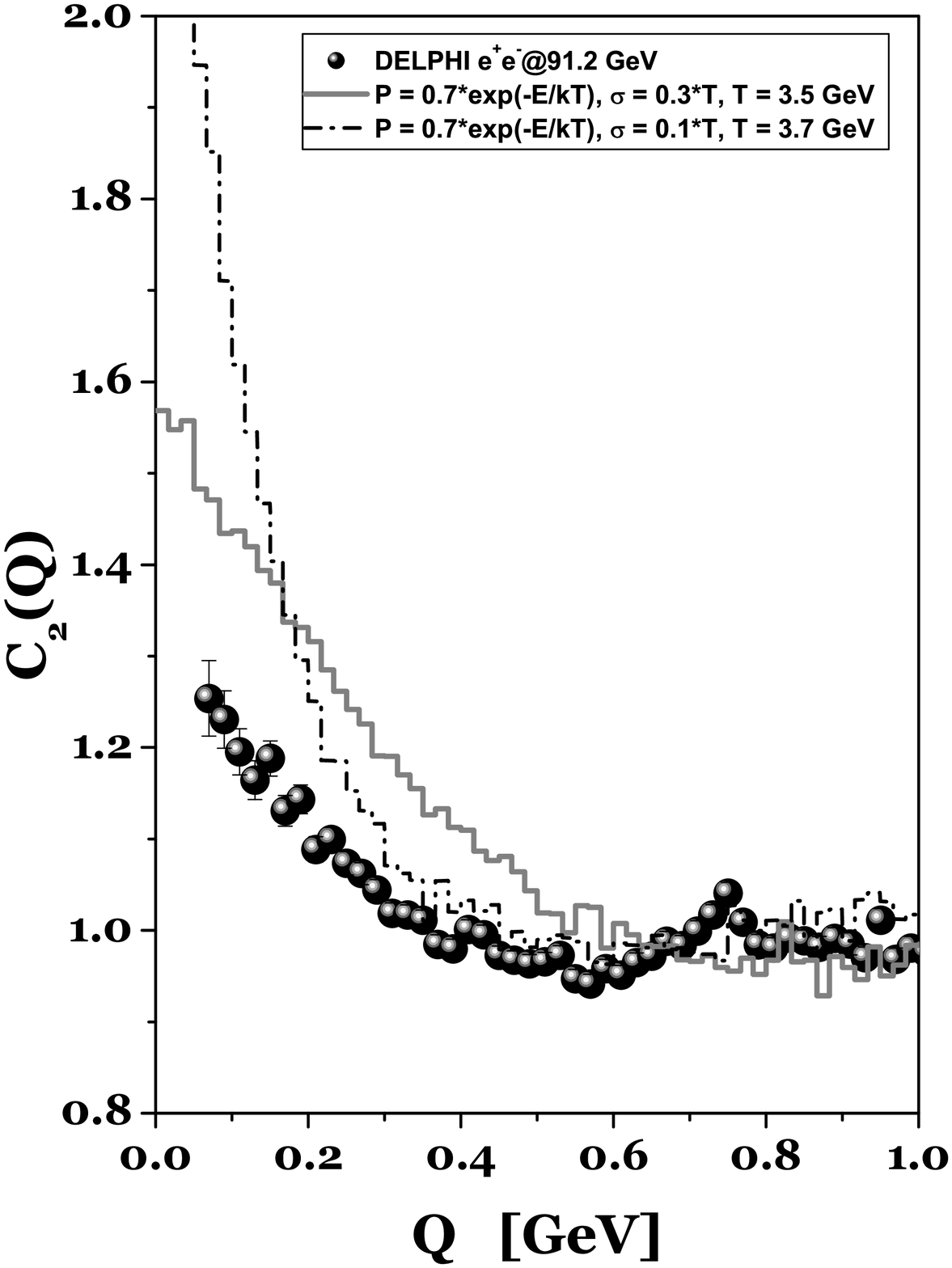}
  \end{minipage}
  \caption{
\footnotesize Examples of confrontation with the experimental data
\protect\cite{Data}. Left panel: fit to charge multiplicity distribution. Right
panel: results for $C_2(Q=|p_1-p_2|)$ correlation function (one dimensional phase
space was used here only). Two different sets of parameters have been used. Notice
that whereas they lead to essentially similar $P(n)$ the resulting $C_2(Q)$ are
drastically different.}
  \label{fig:Fig2}
\end{figure}
\vspace{-5mm}\\ {\bf 3)} To get the experimentally observed width of
$C_2(Q)$ one has to allow that particles in each EEC can have
(slightly) different energies from energy of the particle being its
seed. To do it one must allow that each additional particle selected
in point $2)$ above have energy $E^{(1)}_i$ selected from some
distribution function peaked at $E_1^{(1)}$, $G\left( E^{(1)}_1 -
E^{(1)}_i\right)$, where the width of this distribution, $\sigma$,
is another free parameter.$^b$

{\bf 4)} Repeat points $1)$ - $3)$ as long as there is enough energy
left. Correct in every event for every energy-momentum
nonconservation caused by the selection procedure adopted and assure
that $N^{(+)}=N^{(-)}$.

As a result, in each event we get a number of EEC with particles of
the same charge and (almost) the same energy, i.e. picture closely
resembling classical {\it clans} of \cite{NBD} (with no effects of
statistics imposed, see Fig. \ref{fig:Fig1}). Our clans (containing
now identical bosonic particles subjected to quantum statistics and
therefore named {\it quantum clans}) are distributed in the same way
as the particles being the seeds for EEC, i.e. according to Poisson
distribution. With particles in each clan distributed according to
geometrical distribution they lead therefore to the overall
distribution in our Quantum Clan Model case being of the so called
P\`olya-Aeppli type \cite{PA}. This distribution strongly resembles
the Negative Binomial distributions obtained in the classical clan
model \cite{NBD} where particles in each clans were assumed to
follow {\it logarithmic} distribution instead (with differences
occurring for small multiplicities \cite{PA}). The first preliminary
results presented in Fig. \ref{fig:Fig2} are quite encouraging
(especially when one remembers that so far effects of resonances and
all kind of final state interactions to which $C_2$ is sensitive
were neglected here). It remains now to be checked what two-body BEC
functions for other components of the momentum differences and how
they depend on the EEC parameters: $T$, $P_0$ and $\sigma$. So far
the main outcome is suggestion that EEC's are among the possible
explanations of the BEC effect, in which case BEC would provide us
mainly with their characteristics. This should clear at least some
of many apparently "strange" results obtained from BEC recently
\cite{Strange}.

\section*{Acknowledgment(s)}
Two of us (OU and GW) are grateful for support from the Hungarian Academy
of Science and for the warm hospitality extended to them by organizers of
the {\it $4^{th}$ Budapest Winter School on Heavy Ion Physics}, Dec. 1-3,
2004, Budapest, Hungary. Partial support of the Polish State Committee
for Scientific Research (KBN) (grant 621/E-78/SPUB/CERN/P-03/DZ4/99 (GW))
is acknowledged.

\section*{Notes}
\begin{notes}
\item[a] It means then that from $C_2$ one gets not the size of
the hadronizing source but only the size of the emitting cell, in
\cite{OMT} $R\sim 1/\delta y$, cf. \cite{Z}. In the quantum field
theoretical formulation of BEC this directly corresponds to the
necessity of replacing delta functions in commutator relations by
well-defined peaked functions introducing in this way the same
dimensional scale to be obtained from the fits to data
\cite{Kozlov}. This fact was known even before but without any
phenomenological consequences \cite{Zal}.
\item[b]  It reflects situation encountered
in \cite{Kozlov} where, as we have already mentioned before, the
observed shape of $C_2(Q)$ was coming from the assumed shape of
function replacing Dirac delta function in energy, i.e. introducing
a smearing in energy.
\end{notes}

\vfill\eject
\end{document}